\documentclass[aps,showpacs,twocolumn,superscriptaddress]{revtex4}
\usepackage[dvips]{graphicx}
\usepackage{amsmath}
\usepackage{color}

\begin{document}

\title{Stretch diffusion and heat conduction in 1D nonlinear lattices}

\author{Zhibin Gao}
\affiliation{Center for Phononics and
Thermal Energy Science and School of Physics Science and
Engineering, Tongji University, 200092 Shanghai, People's Republic
of China}

\author{Nianbei Li}
\email{nbli@tongji.edu.cn} \affiliation{Center for Phononics and
Thermal Energy Science and School of Physics Science and
Engineering, Tongji University, 200092 Shanghai, People's Republic
of China}

\author{Baowen Li}
\email{phononics@tongji.edu.cn} \affiliation{Center for Phononics
and Thermal Energy Science and School of Physics Science and
Engineering, Tongji University, 200092 Shanghai, People's Republic
of China} \affiliation{Department of Physics and Centre for
Computational Science and Engineering, National University of
Singapore, Singapore 117546, Republic of Singapore} \affiliation{Center for Advanced 2D Material and Graphene Research Centre, National University of Singapore, Singapore 117542, Republic of Singapore}


\begin{abstract}
In the study of 1D nonlinear Hamiltonian lattices, the conserved quantities play an important role in determining the actual behavior of heat conduction. Besides the total energy, total momentum and total stretch could also be conserved quantities. In microcanonical Hamiltonian dynamics, the total energy is always conserved. It was recently argued by Das and Dhar that whenever stretch (momentum) is not conserved in a 1D model, the momentum (stretch) and energy fields exhibit normal diffusion. In this work, we will systematically investigate the stretch diffusions for typical 1D nonlinear lattices. No clear connection between the conserved quantities and heat conduction can be established. The actual situation is more complicated than what Das and Dhar claimed.
\end{abstract}

\maketitle
\section{Introduction}
The anomalous heat conduction was first predicted for the 1D Fermi-Pasta-Ulam $\beta$ (FPU-$\beta$) nonlinear lattices by Lepri {\it et al} in 1997 \cite{Lepri1997prl}. In this pioneering work, it was found numerically that the thermal conductivity $\kappa$ diverges with the system size $N$ as $\kappa\propto N^{\alpha}$ with $0<\alpha<1$ which breaks the Fourier's heat conduction law \cite{Lepri1997prl}. Numerical simulations also confirm this anomalous heat conduction in diatomic Toda lattice \cite{Hatano1999pre}, carbon nanotubes \cite{Zhang2005jcp} and single polymer chains \cite{LiuJun2012prb}, to name a few. On the other hand, the 1D nonlinear lattices with external on-site potential such as Frenkel-Kontorova (FK) and $\phi^4$ lattices show normal heat conduction \cite{Hu1998pre,Aoki2000pla,Hu2000pre}. Much efforts had been devoted to unraveling the physical mechanism behind normal and anomalous heat conduction in low dimensional systems \cite{Lepri1998epl,Lepri1998pre,Tong1999prb,Tsironis1999pre,Sarmiento1999pre,Dhar1999prl,Alonso1999prl,Li2001prl,Dhar2001prl,Aoki2001prl,Zhang2002pre,Li2002prl,
Narayan2002prl,Denisov2003prl,Saito2003epl,Savin2003pre,Lepri2003pre,Segal2003jcp,Wang2004prl,Gendelman2004prl,Li2005chaos,Dadswell2005pre,Delfini2006pre,Pereira2006prl,Dubi2009pre,Henry2009prb,
Yang2010nt,Wang2011epl,Xiong2012pre,Landi2013pre,Pereira2013pre,Chen2013pre,Xiong2014pre,Wang2015pre,Dadswell2015pre}. The consensus reached in this community is that the momentum conservation and dimensionality play the important roles in determining the actual heat conduction behavior \cite{Lepri2003pr,Dhar2008ap,Liu2013epjb}. The mode coupling theory \cite{Lepri2003pr} predicts that $\kappa\propto N^{\alpha}$, $\kappa\propto \ln{N}$ and $\kappa\propto \mbox{const}$ for 1D, 2D and 3D momentum conserved systems, respectively. The numerics in 2D and 3D lattice systems were found to be consistent with these predictions \cite{Lippi2000jsp,Yang2006pre,Xiong2010pre,Wang2012pre,Saito2010prl,Wang2010prl}. In particular, the prediction of length-dependent anomalous heat conduction were also verified experimentally in 1D nanotubes \cite{Chang2008prl} and molecular chains \cite{Meier2014prl} and 2D suspended graphene \cite{Xu2014nc}. However, there is one exception of 1D coupled rotator lattice which displays normal heat conduction behavior despite its momentum conserving nature \cite{Giardina2000prl,Gendelman2000prl,Yang2005prl,Li2014arxiv}.

The traditional numerical methods used to investigate the heat conduction problem are Non-Equilibrium Molecular Dynamics (NEMD) and equilibrium Green-Kubo (GK) method \cite{Lepri2003pr,Dhar2008ap}. A novel diffusion method in thermal equilibrium was proposed by Zhao \cite{Zhao2006prl} which opens a new way to explore the heat transport problem in nonlinear systems \cite{Li2010prl,Li2014arxiv}. The mean square displacement of energy diffusion generally follows a power-law time dependence as $\left<\Delta x^2(t)\right>_E\propto t^{\beta}$. It has also been rigorously proven \cite{Liu2014prl} that this energy diffusion method is equivalent to the Green-Kubo method where the connection relation of $\alpha=\beta-1$ firstly proposed from particle diffusion analysis \cite{Denisov2003prl} can be derived.

There are also continued theoretical efforts ranging from early mode-coupling theory \cite{Lepri1998epl,Lepri1998pre,Wang2004prl}, renormalization group analysis \cite{Narayan2002prl}, hydrodynamical theory \cite{Dadswell2005pre,Dadswell2015pre}, self-consistent mode-coupling theory \cite{Delfini2006pre}, to recent nonlinear fluctuating hydrodynamical theory \cite{Spohn2014JSP,Beijeren2012prl,Mendl2013prl,Mendl2014pre,Das2014pre,Spohn2014arxiv,Mendl2015arxiv,Spohn2015arxiv}. Although there is still debate about the actual classification and divergent exponents of the universal classes, these theoretical works have greatly improved our understanding on the nature of the anomalous heat transport in low dimensional systems definitely.

Most recently, it was claimed by Das and Dhar that ``whenever stretch (momentum) is not conserved in a one-dimensional model, the momentum (stretch) and energy fields exhibit normal diffusion" \cite{Dhar2015arxiv}. The 1D coupled rotator lattice was taken as the example to support this claim. However, after carefully studying some typical 1D nonlinear lattices with normal heat conduction or energy diffusion behaviors, we found no obvious connection between the stretch and momentum conservation and normal energy and stretch diffusion can be established. Our numerical results indicate that the actual situation might be more complicated than what has been claimed.

This paper will be organized as the following. In Section II, we will present the detailed numerical results of stretch and energy diffusion for typical 1D nonlinear lattices such as $\phi^4$, coupled rotator, FK, combined (FK+$\phi^4$) lattices. The conclusions will be summarized in Section III.

\section{Stretch diffusion in typical 1D nonlinear lattices}
We consider the following Hamiltonian for general 1D lattices
\begin{equation}
H=\sum_i H_i=\sum_i\left[\frac{p^2_i}{2}+V(q_{i+1}-q_i)+U(q_i)\right]
\end{equation}
where $q_i$ and $p_i$ denote the displacement and momentum for the $i$-th atom, respectively. The interaction potential $V(q_{i+1}-q_i)$ only depends on the displacement difference of $(q_{i+1}-q_i)$. The existence of on-site potential $U(q_i)$ will break the conservation of total momentum. For simplicity, periodic boundary conditions $q_i=q_{N+i}$ are applied. The atom index $i$ is assigned as $-(N-1)/2,...,-1,0,1,...,(N-1)/2$ where an odd number of lattice sizes $N$ is chosen without loss of generality.

\begin{figure}
\includegraphics[width=\columnwidth]{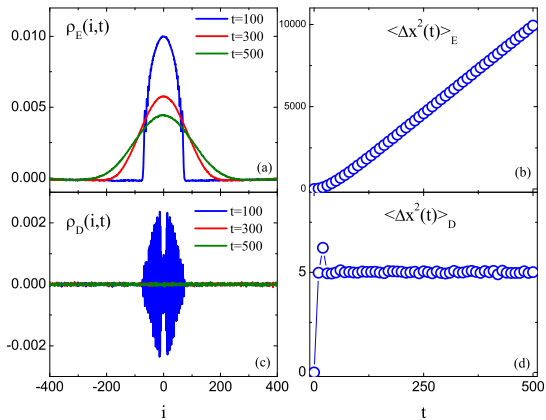}
\vspace{-0.5cm} \caption{\label{fig:phi4_rho_msd}
(color online). The $\phi^4$ lattice. (a) The energy correlation function $\rho_E(i,t)$ and (c) stretch correlation function $\rho_D(i,t)$. (b) The MSD $\left<\Delta x^2(t)\right>_E$ of energy and (d) $\left<\Delta x^2(t)\right>_D$ of stretch. The energy correlation function $\rho_E(i,t)$ follows the Gaussian distribution when correlation time $t>100$ while the stretch correlation function $\rho_D(i,t)$ fails to follow the Gaussian distribution. As a result, the MSD of energy depends on time linearly as $\left<\Delta x^2(t)\right>_E\propto t$, displaying the normal diffusion behavior. In contrast, the MSD $\left<\Delta x^2(t)\right>_D$ of stretch saturates to a constant value after a short time scale. In panel (a), the $\rho_E(i,t)$ is shifted upward with a constant value of $1/(N-1)$ to maintain vanishing tail close to the boundaries \cite{Chen2013pre}. The energy density is set as $E=0.4$ and the corresponding temperature is around $T\approx 0.44$. The lattice size is chosen as $N=801$.}
\end{figure}

In order to study the stretch diffusion and energy diffusion behaviors for 1D lattices, we adopt the definition for the excess energy distribution function $\rho_E(i,t)$ as\cite{Zhao2006prl}
\begin{equation}
\rho_E(i,t)=\frac{\left<\Delta H_i(t)\Delta H_0(0)\right>}{\left<\Delta H_0(0)\Delta H_0(0)\right>}
\end{equation}
where $\Delta H_i(t)=H_i(t)-\left<H_i\right>$ and $\left<\cdot\right>$ denotes the ensemble average or time average equivalently for ergodic systems. The stretch distribution function $\rho_D(i,t)$ can also be defined similarly as\cite{Spohn2014JSP}
\begin{equation}
\rho_D(i,t)=\frac{\left<\Delta D_i(t)\Delta D_0(0)\right>}{\left<\Delta D_0(0)\Delta D_0(0)\right>}
\end{equation}
where the local stretch $D_i(t)\equiv q_{i+1}(t)-q_i(t)$ and $\Delta D_i(t)=D_i(t)-\left<D_i\right>$.

\begin{figure}
\includegraphics[width=\columnwidth]{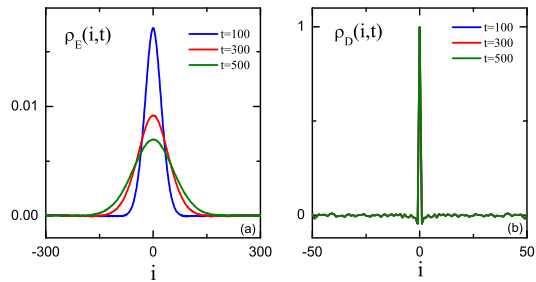}
\vspace{-0.5cm}
\caption{\label{fig:rotatororiginal}
(color online). The coupled rotator lattice of case (i). (a) The energy correlation function $\rho_E(i,t)$ and (b) stretch correlation function $\rho_D(i,t)$. The energy correlation functions $\rho_E(i,t)$ follow the Gaussian distributions at correlation times $t=100,300$ and $500$, implying a normal diffusion behavior. However, the stretch correlation functions $\rho_D(i,t)$ all collapse to the same pattern curve as that at $t=0$.  The energy density is set as $E=0.5$ and the corresponding temperature is around $T\approx 0.54$. The lattice size is chosen as $N=601$.}
\end{figure}

In microcanonical systems with periodic boundary conditions, both the total energy $H=\sum_i H_i$ and total stretch $D=\sum_i D_i=\sum_i (q_{i+1}-q_i)$ are conserved quantities. As a result, the excess energy distribution function $\rho_E(i,t)$ and the stretch distribution function $\rho_D(i,t)$ satisfy the sum rules as $\sum_i\rho_E(i,t)=\sum_i\rho_D(i,t)=0$ in microcanonical systems \cite{Chen2013pre} by noticing that $\sum_i A_i(t)-\sum_i \left<A_i\right>=0$ with $A_i=H_i,D_i$.

One can also define a momentum distribution function $\rho_P(i,t)$ \cite{Zhao2006prl,Li2014arxiv}. However, for lattices where momentum is not conserved, the summation of $\rho_P(i,t)$ is not time invariant as $\sum_i\rho_P(i,t)\neq 0$. In this case, it is meaningless to discuss the momentum diffusion.

In numerical simulations, the fourth order symplectic algorithm will be used to integrate the equations of motions for 1D lattices. The time steps $\Delta t=0.1$ or $0.05$ will be adopted. With this numerical setup, the sum of $\sum_i\rho_E(i,t)$ and $\sum_i\rho_D(i,t)$ can be maintained within the range of the order of $10^{-5}$ and $10^{-14}$, respectively. The energy density $E=H/N$ is the input parameter and the temperature $T\equiv \left<p^2_i\right>$ is a derived quantity as for isolated microcanonical systems.

\begin{figure}
\includegraphics[width=0.8\columnwidth]{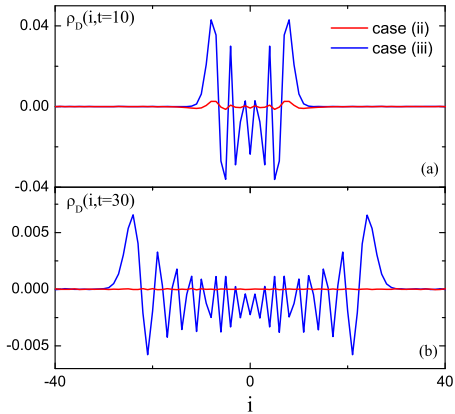}
\vspace{-0.5cm}
\caption{\label{fig:rotator_compare_rhod}
(color online). The stretch correlation functions $\rho_D(i,t)$ for coupled rotator lattice in case (ii) and (iii). At correlation time $t=10$, both $\rho_D(i,t)$ show similar pattern for case (ii) and (iii) except that the amplitude in case (iii) is much larger than that in case (ii). At $t=30$, $\rho_D(i,t)$ still maintains a clear pattern for case (iii) while the pattern disappears for case (ii). The parameters are the same as that used in Fig. \ref{fig:rotatororiginal}.}
\end{figure}

\subsection{$\phi^4$ lattice}
We first consider the 1D $\phi^4$ lattice with the following Hamiltonian
\begin{equation}
H=\sum_i\left[\frac{p^2_i}{2}+\frac{1}{2}(q_{i+1}-q_i)+\frac{1}{4}q^4_i\right]
\end{equation}
The 1D $\phi^4$ lattice is a typical nonlinear lattice with on-site potential which shows normal heat conduction behavior \cite{Hu2000pre,Aoki2000pla}. The total momentum is not conserved due to the existence of external on-site potential. It has been verified that the energy diffusion is normal as well\cite{Zhao2006prl}.

This normal diffusion for energy can be seen from Fig. \ref{fig:phi4_rho_msd} (a) and (b). The excess energy distribution functions $\rho_E(i,t)$ collapse to an almost Gaussian distribution $\rho_E(i,t)\sim \frac{1}{\sqrt{4\pi D_E t}}e^{-\frac{i^2}{4D_E t}}$ at long enough correlation times. As a result, the Mean Square Displacement (MSD) $\left<\Delta x^2(t)\right>_E$ of energy follows a linear time dependence as $\left<\Delta x^2(t)\right>_E\sim 2D_E t$, asymptotically. Here $D_E$ denotes the diffusion constant for energy.

However, the stretch distribution $\rho_D(i,t)$ fails to follow the Gaussian distribution, as can be seen in Fig. \ref{fig:phi4_rho_msd} (c). The two humps at correlation time $t=100$ spreads rapidly over the lattice and disappears at latter correlation times. In Fig. \ref{fig:phi4_rho_msd}, the MSD $\left<\Delta x^2(t)\right>_D$ of stretch is plotted as the function of correlation time $t$. The $\left<\Delta x^2(t)\right>_D$ saturates to a constant value after a short correlation time scale. It is definitely not the normal diffusion behavior which is predicted by Das and Dhar in Ref. \cite{Dhar2015arxiv}. The momentum is not conserved for 1D $\phi^4$ lattice, while its stretch diffusion is not normal!

\begin{figure}
\includegraphics[width=0.8\columnwidth]{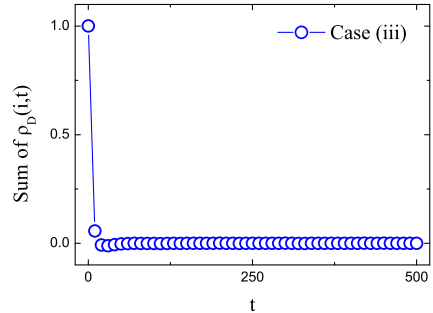}
\vspace{-0.5cm}
\caption{\label{fig:rotator_dhar_sum}
(color online). The sum of $\rho_D(i,t)$ for coupled rotator lattice in case (iii). It can be seen that $\sum_i\rho_D(i,t)$ decays from 1 to 0 very quickly which is consistent with the result in Ref. \cite{Dhar2015arxiv}. As a comparison, the sum $\sum_i\rho_D(i,t)$ for case (ii) can be maintained within the order of $10^{-14}$ in the whole correlation time range. This indicates that the stretch is conserved in case (ii) but not conserved in case (iii). The parameters are the same as that used in Fig. \ref{fig:rotatororiginal}.}
\end{figure}

\subsection{Coupled rotator lattice}
The 1D coupled rotator lattice has the following Hamiltonian
\begin{equation}
H=\sum_i\left[\frac{p^2_i}{2}+[1-\cos{(q_{i+1}-q_i)}]\right]
\end{equation}
Although it conserves the total momentum, it has normal heat conduction behavior \cite{Giardina2000prl,Gendelman2000prl} as well as normal energy diffusion behavior \cite{Li2014arxiv}. In particular, its momentum diffusion is also normal \cite{Li2014arxiv}.

The stretch conservation is a tricky issue for coupled rotator lattice due to the $2\pi$ degeneracy of $q_i$. The dynamics of the system is invariant to the arbitrary shift of multiple $2\pi$ for every $q_i$ as $q_i\rightarrow q_i+2n\pi$, where $n$ can be any integer number. Depending on how to limit the $q_i$, the stretch of coupled rotator lattice can be a conserved quantity or not. We consider the following three limitations for $q_i$ or $D_i$.

(i) No limitations. Nothing is done to the value of $q_i$. The $q_i$ can take whatever it takes during the evolution of the systems. In this situation, the stretch is a conserved quantity. The local stretch $D_i=q_{i+1}-q_i$ can take value from negative infinity to positive infinity and the partition function is not well defined \cite{Dhar2015arxiv,Spohn2014arxiv}. Although this effect will cause problem in theoretical analysis, the dynamics of the system will not be affected. The energy diffusion is normal as can be seen from Fig. \ref{fig:rotatororiginal} (a). In this situation, the stretch correlation functions $\rho_D(i,t)$ all collapse to the same pattern as that at $t=0$ in Fig. \ref{fig:rotatororiginal} (b). This might be due to the unbounded fact of the value of $q_i$.

\begin{figure}
\includegraphics[width=\columnwidth]{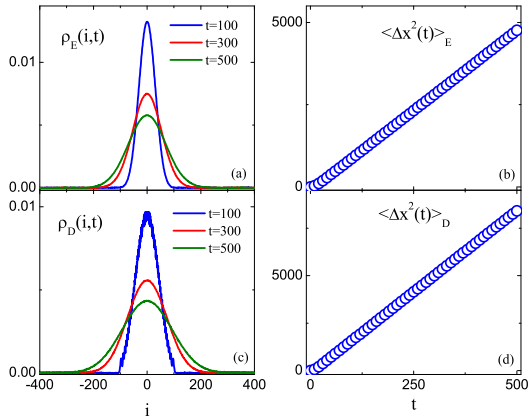}
\vspace{-0.5cm} \caption{\label{fig:fk_rho_msd}
(color online). The FK lattice. (a) The energy correlation function $\rho_E(i,t)$ and (c) stretch correlation function $\rho_D(i,t)$. (b) The MSD $\left<\Delta x^2(t)\right>_E$ of energy and (d) $\left<\Delta x^2(t)\right>_D$ of stretch. Both the energy correlation function $\rho_E(i,t)$ and the stretch correlation function $\rho_D(i,t)$ follow the Gaussian distribution after long enough correlation time. As a result, both the energy and stretch diffusion is normal as $\left<\Delta x^2(t)\right>_{E/D}\propto t$ asymptotically. In panel (a) and (c), the $\rho_{E/D}(i,t)$ are shifted upward with a constant value of $1/(N-1)$ to maintain vanishing tail close to the boundaries. The parameter for on-site coupling strength is set as $V=1$. The energy density is set as $E=1$ and the corresponding temperature is around $T\approx 0.86$. The lattice size is chosen as $N=801$.}
\end{figure}

(ii) The $q_i$ is limited within $-\pi<q_i\le \pi$. After each time step in numerical simulations, the $q_i$ is forced to reshifted into this region whenever it jumps out. As a result, the local stretch $D_i$ lies within $-2\pi<D_i\le 2\pi$. The stretch is a still conserved quantity and the partition function is well defined. In numerical simulations, the sum $\sum_i\rho_D(i,t)$ can be maintained within the order of $10^{-14}$ for all times which is a signature of conserved quantity. The $\rho_D(i,t)$ displays similar spatial pattern at short correlation times as that in case (iii). The only difference is that the amplitude is much smaller in case (ii), as can be seen from Fig. \ref{fig:rotator_compare_rhod} (a). At larger correlation times see Fig. \ref{fig:rotator_compare_rhod} (b), the $\rho_D(i,t)$ quickly loses its spatial pattern in comparison to that in case (iii). The MSD $\left<\Delta x^2(t)\right>_D$ of stretch in this case saturates to a constant value after a short correlation time (not shown here), similar to that of $\phi^4$ lattice.

(iii) The local stretch $D_i$ is limited within $-\pi<D_i\le \pi$. The $q_i$ is not affected during the dynamical evolution. However, $D_i$ is adjusted appropriately every time when it is recorded to generate the correlation function. Therefore, the partition function is well defined. Only in this special situation, the stretch is not a conserved quantity as can be seen from Fig. \ref{fig:rotator_dhar_sum}.

From the above results and discussions, it is found that the conservation of stretch is a very tricky issue. Depends on the limitation of $q_i$ or $D_i$, the stretch can be adjusted to be a conserved or nonconserved quantity.

\begin{figure}
\includegraphics[width=\columnwidth]{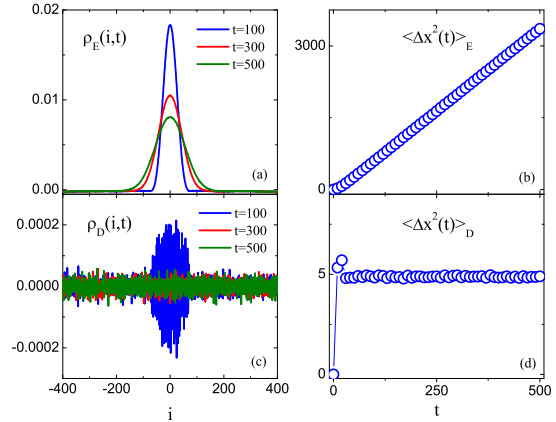}
\vspace{-0.5cm}
\caption{\label{fig:fkphi4_rho_msd}
(color online). The combined FK+$\phi^4$ lattice. (a) The energy correlation function $\rho_E(i,t)$ and (c) stretch correlation function $\rho_D(i,t)$. (b) The MSD $\left<\Delta x^2(t)\right>_E$ of energy and (d) $\left<\Delta x^2(t)\right>_D$ of stretch. The overall behavior is similar to the $\phi^4$ lattice. Only the energy correlation function $\rho_E(i,t)$ follows the Gaussian distribution. As a result, only the energy diffusion is normal as $\left<\Delta x^2(t)\right>_{E}\propto t$ asymptotically. In panel (a), the $\rho_E(i,t)$ is shifted upward with a constant value of $1/(N-1)$ to maintain vanishing tail close to the boundaries. The parameter is set as $V=0.5$. The energy density is set as $E=0.5$ and the corresponding temperature is around $T\approx 0.47$. The lattice size is chosen as $N=801$.}
\end{figure}

\subsection{FK lattice}
We then consider another nonlinear lattice with on-site potential, the FK lattice with Hamiltonian as
\begin{equation}\label{ham-fk}
H=\sum_i\left[\frac{p^2_i}{2}+\frac{1}{2}(q_{i+1}-q_i)^2+\frac{V}{2\pi}(1-\cos{2\pi q_i})\right]
\end{equation}
The FK lattice also exhibits normal heat conduction \cite{Hu1998pre} as well as normal energy diffusion behaviors. In Fig. \ref{fig:fk_rho_msd} (a) and (b), the excess energy distribution function $\rho_E(i,t)$ and the MSD $\left<\Delta x^2(t)\right>_E$ of energy are plotted. The $\rho_E(i,t)$ follows the Gaussian distribution functions and the $\left<\Delta x^2(t)\right>_E$ is linearly proportional to the correlation time as $\left<\Delta x^2(t)\right>_E\propto t$, indicating obvious normal diffusion behavior for energy.

In contrast to the $\phi^4$ lattice, the stretch distribution function $\rho_D(i,t)$ approaches the Gaussian distributions as $\rho_D(i,t)\sim \frac{1}{\sqrt{4\pi D_D t}}e^{-\frac{i^2}{4D_D t}}$ at long enough correlation times for the FK lattice, as can be seen in Fig. \ref{fig:fk_rho_msd} (c). The MSD $\left<\Delta x^2(t)\right>_D$ of stretch do follow the linear time dependence as $\left<\Delta x^2(t)\right>_D\sim 2D_D t$ asymptotically in in Fig. \ref{fig:fk_rho_msd} (d). The stretch diffusion is normal for FK lattice despite its stretch conservation nature!

Unlike the coupled rotator lattice, there is no ambiguous space for the stretch conservation in FK lattice. The $q_i$ or $D_i$ is not $2\pi$ invariant anymore due to the existence of the interaction potential in Eq. (\ref{ham-fk}).

\subsection{Combined (FK+$\phi^4$) lattice}
In the end, we consider the combined (FK+$\phi^4$) lattice with Hamiltonian
\begin{equation}
H=\sum_i\left[\frac{p^2_i}{2}+\frac{1}{2}(q_{i+1}-q_i)^2+\frac{V}{2\pi}(1-\cos{2\pi q_i})+\frac{1}{4}q^4_i\right]
\end{equation}
The combined FK+$\phi^4$ lattice should also have normal heat conduction behavior. This can be verified by examining the energy diffusion behavior in Fig. \ref{fig:fkphi4_rho_msd} (a) and (b). The $\rho_E(i,t)$ follows the Gaussian distributions and the MSD of energy depends linearly on time as $\left<\Delta x^2(t)\right>_E\sim 2D_E t$ asymptotically.

In Fig. \ref{fig:fkphi4_rho_msd} (c), the stretch distribution functions $\rho_D(i,t)$ are plotted for correlation times $t=100,300$ and $500$. No Gaussian behavior is observed and the scenario is similar to that of $\phi^4$ lattice as in Fig. \ref{fig:phi4_rho_msd} (c). Same as $\phi^4$ lattice, the $\left<\Delta x^2(t)\right>_D$ saturates to a constant value after a short time scale, as can be seen in Fig. \ref{fig:fkphi4_rho_msd} (d). This is another example that the momentum is not conserved, while the stretch diffusion is not normal.

\section{Conclusions}
In conclusion, we have systematically investigated the stretch diffusion as well as the energy diffusion for a few 1D nonlinear lattices with normal heat conduction behaviors. For isolated lattices with periodic boundary conditions, both the total energy and total stretch are conserved quantities. Depends on the existence of on-site potential, the total momentum can be conserved or nonconserved. For $\phi^4$ and combined (FK+$\phi^4$) lattices, the total momentum is not conserved while the stretch diffusion is not normal. For FK lattice, the total momentum is not conserved and the stretch diffusion is normal. For coupled rotator lattice with normal momentum diffusion, the situation is tricky in the sense that its stretch conservation depends on the choices of limitation of $q_i$ or $D_i$. Our numerical results do not support the claim that``whenever stretch (momentum) is not conserved in a one-dimensional model, the momentum (stretch) and energy fields exhibit normal diffusion" proposed by Das and Dhar in Ref. \cite{Dhar2015arxiv}. However, there is something interesting for the lattices with cosine or bounded potentials. It is still an open issue and we hope more efforts will be done in this direction in the future.

\section{acknowledgments}
The numerical calculations were carried out at Shanghai Supercomputer Center, which
has been supported by the NSF China with grant No. 11334007(B.L.). This work has
been supported by the NSF China with grant No. 11334007(Z.G., N.L., B.L.), the NSF
China with Grant No. 11205114(N.L.), the Program for New Century Excellent Talents
of the Ministry of Education of China with Grant No. NCET-12-0409(N.L.) and the
Shanghai Rising-Star Program with grant No. 13QA1403600(N.L.).

\bibliographystyle{apsrev4-1}

\end{document}